\documentclass[12pt,preprint]{aastex}

\begin{document}

\slugcomment{To appear in the Astrophysical Journal March 20, 2005}

\title{Observations and modeling of the inner disk region of
T Tauri stars}

\author{R.L. Akeson\altaffilmark{1}, C.H. Walker\altaffilmark{2},
K. Wood\altaffilmark{2}, J.A. Eisner\altaffilmark{3},
E. Scire\altaffilmark{4}, B. Penprase\altaffilmark{4},
D.R. Ciardi\altaffilmark{1}, G.T. van~Belle\altaffilmark{1},
B. Whitney\altaffilmark{5}, and J.E. Bjorkman\altaffilmark{6} }

\altaffiltext{1}{Michelson Science Center, California Institute of Technology
, MS 100-22, Pasadena, CA, 91125}

\altaffiltext{2}{School of Physics and Astronomy, University of St. Andrews, North Haugh, St. Andrews, KY16 9AD, Scotland}

\altaffiltext{3}{Dept. of Astronomy, California Institute of Technology,
MS 105-24, Pasadena, CA, 91125}

\altaffiltext{4}{Department of Physics and Astronomy, Pomona College, Claremont, CA 91711}

\altaffiltext{5}{Space Science Institute, 3100 Marine Street, Suite A353, Boulder, CO 80303}

\altaffiltext{6}{Ritter Observatory, Department of Physics and Astronomy, University of Toledo, Toledo, OH 43606}

\begin{abstract}
We present observations of four T Tauri stars using long baseline
infrared interferometry from the Palomar Testbed Interferometer.  The
target sources, T Tau N, SU Aur, RY Tau and DR Tau, are all known to
be surrounded by dusty circumstellar disks.  The observations directly
trace the inner regions ($<$1 AU) of the disk and can be used to
constrain the physical properties of this material.  For three of the
sources observed, the infrared emission is clearly resolved.  We first
use geometric models to characterize the emission region size, which
ranges from 0.04 to 0.3 AU in radius.  We then use Monte Carlo
radiation transfer models of accretion disks to jointly model the
spectral energy distribution and the interferometric observations with
disk models including accretion and scattering.  With these models, we
are able to reproduce the data set with extended emission arising from
structures larger than 10 milliarcseconds contributing less than 6\%
of the K band emission, consistent with there being little or no
envelope remaining for these Class II sources ($d \log (\lambda
F_{\lambda})/d \log \lambda \approx $ -2 to 0 in the infrared).  The
radiation transfer models have inner radii for the dust similar to the
geometric models; however, for RY Tau emission from gas within the
inner dust radius contributes significantly to the model flux and
visibility at infrared wavelengths.  The main conclusion of our
modeling is that emission from inner gas disks (between the magnetic
truncation radius and the dust destruction radius) can be a
significant component in the inner disk flux for sources with large
inner dust radii.

\end{abstract}


\section{Introduction}

In the canonical model, T Tauri systems comprise the central star, a
rotating disk of gas and dust, a jet or outflow and possibly a
residual circumstellar envelope (see e.g. \citet{ber89}).  In many
cases, the central star is still accreting material and this process,
as well as the mechanisms driving the outflow, are dependent on and
influence the properties of the inner disk ($<$1 AU).  Several groups
(e.g. \citet{kon91} and \citet{shu94}) have proposed models in which
the stellar magnetic field truncates the disk at a few stellar radii.
Matter from the disk flows along the field lines and onto the star
producing hot spots or rings that can explain observed ultraviolet
photometric variability \citep{ken94,woo96,gom97}.

In the last several years, the technique of long-baseline infrared
interferometry has been applied to the study of circumstellar material
around young stellar objects.  These observations are sensitive to hot
material near the star itself.  Given the milliarcsecond resolution
capability of the current generation of interferometers, these
observations can in many cases spatially resolve the emission from the
hot (a few thousand Kelvin) material and are well suited for
observations of the inner regions of young stellar objects.  The first
young stellar object to be observed using this technique was FU Ori
\citep{mal98}, followed by Herbig Ae/Be stars \citep{mil99,mil01} and
T Tauri stars \citep{ake00}(hereafter Paper 1).  The FU Ori results
were consistent with accretion disk models, while both the T Tauri and
Herbig star results found characteristic sizes larger than expected
from geometrically flat accretion disk models.  More recent
observations of Herbigs \citep{eis04} have found earlier spectral type
objects which are consistent with accretion disk predictions.

Measurements of the spectral energy distribution (SED) at optical
through radio wavelengths probe a range of processes in young stellar
objects including the stellar photosphere, accretion onto the star or
disk, emission from gas and dust in the disk and emission from the
outflow.  In many sources, continuum emission from circumstellar or
accreting material adds to the stellar spectrum, decreasing the
stellar spectral features in an effect called veiling.  For T Tauri
stars, the veiling in the infrared can very high, indicating
substantial excess emission (see e.g \citet{fol99}).

In Paper 1 we presented observations showing that the infrared
emission from the T Tauri stars T~Tau~N and SU~Aur is resolved.  The
visibilities from T Tauri stars can be difficult to model given the
substantial stellar component, infrared variability and the possible
presence of a significant extended component.  In this paper, we
present further interferometric observations of the T Tauri stars T
Tau N, SU Aur, DR Tau and RY Tau using the Palomar Testbed
Interferometer (PTI) and infrared photometry from the Pomona College 1-meter
telescope.  In \S \ref{model}, we present geometric models to constrain
the emission size and orientation.  In \S \ref{scatter}, we present
detailed source models which include the scattered light and
reprocessing of starlight and dissipation of viscous accretion
energy in the disk to fit both the SED and the infrared visibilities.

\section{Sources}

All four sources are located in the Taurus-Auriga molecular cloud
(distance $\sim$ 140~pc) and are well studied T Tauri objects.  Source
spectral types and stellar properties given in Table \ref{table:source}
are taken from recent references using infrared spectroscopy.
Due to the sensitivity restrictions of PTI, we have chosen sources which are
among the most infrared luminous T Tauri objects.  As the PTI
acquisition system works in the optical, there is a selection
effect against highly inclined, optically
obscured sources.

\begin{table}[h!]
\begin{center}
\begin{tabular}{lclll} \hline
Source & Sp Type & L$_{\star}$ (L$_{\odot}$) & R$_{\star}$ (R$_{\odot}$) & Ref. \\ \hline
T Tau N & K0 & 7.3 & 2.8 & \citet{whi01}\\
SU Aur & G2 & 12.9 & 3.5 & \citet{muz03}\\
DR Tau & K7 & 0.87 & 1.9 & \citet{muz03}\\ 
RY Tau & K1 & 12.8 & 3.6 & \citet{muz03}\\ \hline
\end{tabular}
\caption{Stellar parameters for the observed sources.  
\label{table:source}}
\end{center}
\end{table}

All four systems have significant emission in excess of the stellar
photosphere from near infrared through millimeter wavelengths and all
are believed to have circumstellar disks.  The T Tau system comprises
the optically visible star T Tau N and its infrared companion T Tau S,
which is itself a binary \citep{kor00}.  The PTI observations are of T
Tau N, the component which dominates the millimeter emission
\citep{ake98}.  SU~Aur has an SED similar to that of T~Tau~N, although
\citet{her88} classified SU~Aur separately from other T~Tauri's due to
its high luminosity and broad absorption lines.  RY Tau is associated
with a reflection nebulosity \citep{nak95} and has millimeter-wave
molecular line emission consistent with a Keplerian disk
\citep{koe95}.  DR Tau is one of the most heavily veiled T Tauri stars
and is highly variable in the optical \citep{gul00} and near-infrared
\citep{ken94}.

\section{\bf Observations}

\subsection{Infrared interferometry}
\label{observations}

Infrared interferometry data were taken at the Palomar Testbed
Interferometer (PTI), which is described in detail by \citet{col99}.
PTI is a long-baseline, direct detection interferometer which utilizes
active fringe tracking in the infrared.  Data presented here were
obtained in the K band (2.2 $\mu$m) in all three PTI baselines: NS
(110 meter), NW (85 meter) and SW (85 meters).  In our analysis below,
we also use the SU~Aur observations described in \citet{ake02} and
Paper 1.  A summary of the new observations is given in Table
\ref{table:obs}.  These data were acquired over a period from 24
September 2001 to 16 October 2003.  The data in the NS and NW
baselines were taken with a 20 millisecond fringe integration time,
while the SW data were taken with a 50 millisecond time, providing
better SNR for these data.

\begin{table}[h!]
\begin{center}
\begin{tabular}{lllllll} \hline
\multicolumn{7}{c}{Observations} \\ \hline
& \multicolumn{2}{c}{NS} &  \multicolumn{2}{c}{NW} &  \multicolumn{2}{c}{SW} \\
& nights & ints & nights & ints & nights & ints  \\
T Tau N & & & & & 1 & 6 \\
SU Aur & & & & & 1 & 6 \\
DR Tau & 3&5 &1 &3 & 1 & 4 \\
RY Tau & 4&27 &3 &14 & 2 & 8 \\ \hline
\multicolumn{7}{c}{Calibrators} \\ \hline
Calibrator & size est.(mas) & Sources\\
HD 28024 & 0.68 & \multicolumn{5}{l}{T Tau N, SU Aur, DR Tau, RY Tau} \\
HD 30111 & 0.60 & \multicolumn{5}{l}{T Tau N, SU Aur} \\
HD 30122 & 0.11 & \multicolumn{5}{l}{T Tau N, SU Aur} \\
HD 28677 & 0.34 & \multicolumn{5}{l}{DR Tau, RY Tau} \\
HD 26737 & 0.24 & \multicolumn{5}{l}{DR Tau, RY Tau} \\ \hline
\end{tabular}
\caption{New observations of T Tauri sources from PTI.  Each integration represents 125 seconds of fringe data. 
\label{table:obs}}
\end{center}
\end{table}

The data were calibrated using the standard PTI method \citep{bod98}.
Briefly, a synthetic wide-band channel is formed from five
spectrometer channels ($\lambda=2.0-2.4$).  The system visibility, the
response of the interferometer to an unresolved object, is measured
using calibrator stars.  The calibrator star sizes were estimated
using a blackbody fit to photometric data from the literature and were
checked to be internally consistent.  Calibrators were chosen for
their proximity to the source and for small angular size, minimizing
systematic errors in deriving the system visibility.  All calibrators
used here have angular diameters $<0.7$ milliarcsecond (mas) and were
assigned uncertainties of 0.1 mas (Table \ref{table:obs}).  The
calibrated data are presented in normalized squared visibility
(V$^2$=1 for an unresolved source), which we refer to as visibility in
this paper.  The calibrated visibility uncertainties are a combination
of the calibrator size uncertainty and the internal scatter in the
data.  As DR~Tau is near the tracking limit for PTI, the wide-band
data are used rather than the synthetic wide-band (spectral channel)
data. The main difference between these two channels is that the
spectral channels are spatially filtered and the wide-band channel is
not. The accuracy of the wide-band data were confirmed by comparing
the wide-band and synthetic wide-band data for other sources observed
on the same night as DR~Tau.

The calibrated data were edited to remove integrations with
very high jitter (a measure of the phase noise) and integrations for which the
estimates of the system visibility from separate calibrator
observations disagreed by more than 3$\sigma$.  In general,
the points eliminated were from entire nights with marginal
weather or integrations taken at large hour angles.  No more than 10\%
of the data for any given source was removed, except for DR Tau on the NS baseline, and inclusion
of these data points would not substantially change the
results given below.

\begin{figure}[h!]
\begin{center}
\epsscale{0.85}
\plotone{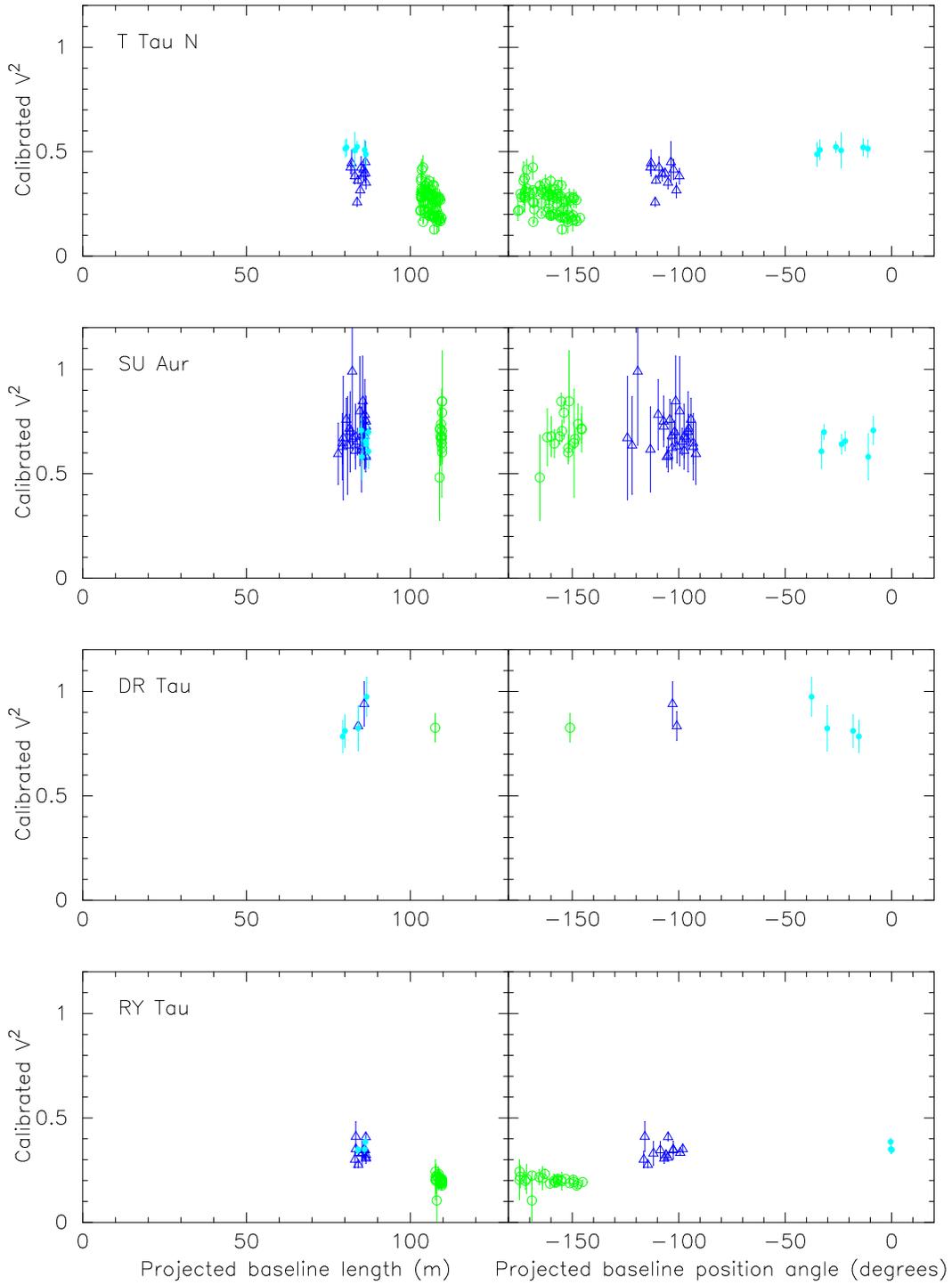}
\caption{Calibrated PTI visibilities for each of the four sources
by baseline: NS (open circles), NW (open triangles) and SW (closed circles).
For T Tau N and SU Aur the data from \citet{ake02} is also
plotted.
\label{fig:data}}
\end{center}
\end{figure}

The calibrated and edited data are shown in Figure \ref{fig:data} for
each source as a function of projected baseline length and position
angle.  Three of the four sources, T~Tau~N, SU~Aur and RY~Tau are
clearly resolved.  The new observations of T Tau N are
consistent with the results of \citet{ake02}.  We have not calculated
models for T~Tau~N here; scattered light models of T~Tau~N that
reproduce the observed asymmetry \citep{sta98} are detailed in
\cite{woo01}.

\subsection{Infrared photometry}

A sample of young stellar objects, including DR Tau, SU Aur and RY Tau
and a sequence of photometric standard stars taken from \citet{lan92}
were observed over seven nights from December 2003 to March 2004
(December 16, January 10, 13, 15, 22, 23, and March 9) using the
Pomona College 1-m telescope with the CLIRCAM infrared camera in the J
and K bands.  For each object, a series of at least five dithered
exposures was used to create individual sky images for each field. The
sky background and instrumental noise was subtracted from all the
images, and repeat exposures were median combined after shifting to a
common astrometric reference frame to remove a majority of the
background noise.  Instrumental magnitudes were converted to standard
J and K magnitudes using a combination of the published magnitudes
and the J and K magnitudes from bright 2MASS stars in
the image frames. The magnitudes given in Table \ref{IRphot:table} are
the average magnitudes over all six nights using the average
calibration zero points from the complete sample of photometric
standards and 2MASS stars.  The three sources observed showed no
statistically significant variability over the nights observed, and
were constant in magnitude within the photometric error of 0.15
magnitudes in J and 0.15 magnitudes in K. For comparison, the 2MASS J
and K magnitude and observation date are also given.  Together, the
Pomona and 2MASS data bracket the PTI observations.  At K band, only
RY Tau shows a significant difference between the 2MASS measurement
and our more recent observations; however, past observations of these
sources have shown infrared variability (particularly DR Tau;
\citet{ken94}).  Additional information on the infrared observations
from Table \ref{IRphot:table} is presented in \citet{pen04}.

\begin{table}[h!]
\begin{center}
\begin{tabular}{lllllll}
Source & \# nights & mag & rms & 2MASS mag & 2MASS rms & 2MASS date \\ \hline
\multicolumn{7}{c}{J band} \\ \hline
SU Aur & 6 & 7.24 & 0.143 & 7.20 & 0.020 & 1/30/98 \\ 
DR Tau & 6 & 8.75 & 0.195 & 8.84 & 0.024 & 10/10/97 \\
RY Tau & 5 & 7.52 & 0.226 & 7.15 & 0.019 & 10/29/97 \\ \hline
\multicolumn{7}{c}{K band} \\ \hline
SU Aur & 6 & 6.17 & 0.114 & 5.99 & 0.022 & 1/30/98 \\ 
DR Tau & 6 & 6.87 & 0.183 & 6.87 & 0.017 & 10/10/97 \\
RY Tau & 6 & 5.76 & 0.168 & 5.39 & 0.022 & 10/29/97 \\ \hline
\end{tabular}
\caption{Results of infrared photometry observations.
\label{IRphot:table}}
\end{center}
\end{table}

\section{Geometric models}
\label{model}

In this section, we discuss geometric models for SU Aur, RY Tau and DR
Tau.  As PTI is a direct detection interferometer, any emission within
the 1\arcsec\ Gaussian (FWHM) field of view will contribute to the
measured visibility.  As discussed in \citet{ake02} there are many
scenarios that could produce a visibility of less than 1.  These
include additional point sources within the field of view, a resolved
source of emission, or extended (over-resolved) emission which will
contribute incoherently.  Any possible incoherent contribution (in
this case any emission on scales greater than 10 mas and within the
1\arcsec\ FOV) is hard to assess for our sources, given that many
observations of envelopes or reflection nebulae do not include the
central arcsecond due to contamination from the star itself.  None of
these three sources has a known companion within 1\arcsec.  DR~Tau was
included in lunar occultation observations of \citet{sim99} and no
detection was reported with a point source limiting magnitude of
$\Delta K$ = 2.5.  Models including scattered light contributions are
presented in \S \ref{scatter} and extended components are discussed in
\S \ref{extended}.

For the model fitting in this section, we adopt a configuration of an
unresolved point source (these stars have diameters $\leq$ 0.1 mas and
therefore a V$^2>0.99$ at PTI) and a resolved component.  We take the
contribution of the stellar component from measurements of the
infrared veiling.  
For SU~Aur and DR~Tau we
use the K band veiling measurements of \citet{muz03}.  For RY~Tau the
\citet{muz03} value of $r_{K}= 0.8 \pm 0.3$ (where $r_K$ = F$_{\rm
excess}$/F$_{\rm star}$) is much less than the lower limit of $r_K > 2.5$
from \citet{fol99}.  For our adopted model (point source + resolved
component), the PTI data and a value of $r_{K}= 0.8$ are incompatible
(i.e. the point source contribution can not be that large and still
produce the PTI measurement) and we therefore use $r_{K}= 2.5 \pm 1$ for RY
Tau in the geometric fits.

Simple geometric models of the emission are used to characterize the
source size and inclination.  The two models presented here use a
uniform disk and a thin ring to represent the emission profile.  For
the uniform disk, only the measured visibility was used to determine
the disk radius.  For the ring model, visibilities were calculated for
a range of inner diameters and compared to the observed visibilities.
For each ring diameter considered, the width was determined by matching
the excess flux, derived using the measured K-band veiling, with a
blackbody emission source at a temperature of 1600~K, the assumed dust
destruction temperature \citep{dus96}.  In these models, the dust
destruction temperature controls the width of the ring, but affects
the fit radius only through the shape of the model visibility curve.
For example, changing the blackbody temperature of the ring from 1200
to 2000 K would change the fit radius for RY Tau by 30\%.  Both
face-on and inclined geometries were fit to the data (Table
\ref{table:fits}).  The uncertainties in the model fits due to the uncertainty
in the stellar contribution are also given.

\begin{table}[h!]
\begin{center}
\begin{tabular}{llll} \hline
& SU Aur & DR Tau & RY Tau \\ \hline 
f$_{\rm excess}$\tablenotemark{a} & $0.44 \pm 0.09$ & 0.8 $\pm 0.3$ & $0.71 \pm 0.11$ \\
K$_m$ (2MASS) & 5.99 & 6.87 & 5.40 \\     \hline
\multicolumn{4}{c}{Face-on models} \\ \hline
\multicolumn{2}{l}{Uniform disk} \\
\quad Radius (AU) & 0.20$\pm 0.028$ & 0.10$\pm 0.029$ & 0.29$\pm 0.036$ \\
\quad $\sigma_v$ (AU)& 0.042 & 0.004 & 0.080 \\
\quad $\chi^2/{dof}$ & 2.5 & 0.85 & 2.9 \\ 
\multicolumn{2}{l}{Ring} \\
\quad Inner radius (AU) & 0.13$\pm 0.021$ & 0.057$\pm 0.027$ & 0.17$\pm 0.01$ \\
\quad Width (AU) & 0.050 & 0.028 & 0.035 \\
\quad $\sigma_v$ (AU) & 0.036 & 0.010 & 0.059 \\
\quad $\chi^2/{dof}$ & 2.5 & 0.85 & 4.6 \\ \hline 
\multicolumn{4}{c}{Inclined models} \\ \hline
\multicolumn{2}{l}{Uniform disk} \\
\quad Radius (AU) & 0.27$\pm 0.037$ & 0.11$\pm 0.03$ & 0.30$\pm 0.008$ \\
\quad PA (degr) & 112 $\pm 24$ & 160 $\pm 55$ & 98 $\pm 40$ \\
\quad Incl (degr) & 51 $\pm 11$ & 40 $\pm 30$ & 19 $\pm 6$ \\
\quad $\chi^2/{dof}$ & 0.9 & 0.77 & 2.3 \\ 
\multicolumn{2}{l}{Ring} \\
\quad Inner radius (AU) & 0.18 $\pm 0.025$ & 0.070 $\pm 0.026$ & 0.19$\pm 0.01$ \\
\quad Width (AU) & 0.008 & 0.019 & 0.029 \\
\quad PA (degr) & 114 $\pm 23$ & 160$\pm 55$ & 110$\pm 22$ \\
\quad Incl (degr) & 52 $\pm 10$ & 40 $\pm 30$ & 25 $\pm$ 3\\
\quad $\chi^2/{dof}$ & 0.9 & 0.78 & 3.0 \\ \hline
\end{tabular}
\caption{Results from geometric model fits.  The systematic error, 
$\sigma_v$ is from the uncertainty in the stellar contribution.
\label{table:fits}}
\tablenotetext{a}{f$_{\rm excess}$ = F$_{\rm excess}$/F$_{\rm total}$, where F$_{\rm total}$ = F$_{\rm star}$ + F$_{\rm excess}$}
\end{center}
\end{table}

\begin{figure}[h!]
\begin{center}
\epsscale{0.4}
\plotone{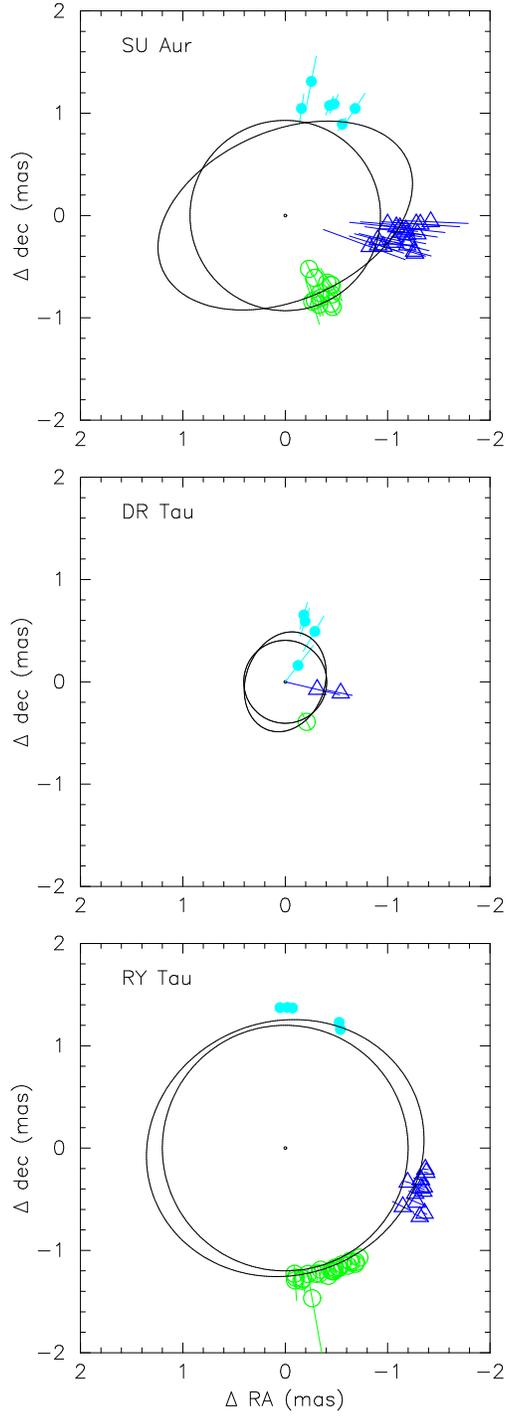}
\caption{The data and uniform disk fits for the best fit face-on
and inclined models.  An unresolved stellar component is included
as described in \S \ref{model}.
Separate symbols are used for each baseline: NS (open circles), NW (open triangles) and SW (closed circles).
\label{fig:UDpoint}}
\end{center}
\end{figure}

The ring model fits are graphically shown in Figure
\ref{fig:UDpoint}.  In this sky plane representation, the radial
coordinate for each data point is the inner ring size corresponding to
the measured visibility and accounting for the stellar component
listed in Table \ref{table:fits}.  The error bars include the errors
on the data points but not the uncertainty in the stellar
contribution.  The polar coordinate is determined by the projected
baseline position angle.  In this way, the constraint provided by the
data on both the size and the inclination are visible.  The best fit
face-on and inclined ring models are also plotted.

\subsection{Discussion}

Using the simple geometric models, we find source sizes ranging from
0.04 to 0.3 AU in radius.  As discussed in Paper 1 and by
\citet{mil01}, the measured sizes for T Tauri stars and Herbig Ae
stars were larger than expected from simple disk models.  An
explanation for this discrepancy in Herbig stars was proposed
independently by two groups based on SED modeling \citep{nat01} and
aperture masking observations \citep{tut01}.  In these models the
inner edge of the dust disk is located at the radius where the dust
reaches the sublimation temperature (R$_{\rm dust}$).  This
configuration produces a vertically extended inner wall, reproducing
both the SED and the interferometry observations for the Herbig
sources.  \citet{dul01} also applied this model to T Tauri stars.
Further work by \citet{muz03} extended the model to include the
accretion luminosity as well as the stellar luminosity in determining
the dust destruction radius for several T Tauri stars, including the
three shown in Figure \ref{fig:UDpoint}.  In all these models,
optically thin gas may be present within R$_{\rm dust}$ (we discuss
this point further in \S \ref{results}).

We chose a ring distribution as a simple representation of a model in
which the infrared emission arises from the inner wall of the dust
disk.  The values for R$_{\rm dust}$ predicted by \citet{muz03} are
larger by roughly a factor of 2 than our fit ring radii.  We note that
the presence of extended emission which was not included in our model
would decrease the fit radius, and would therefore increase this
discrepancy.  The fit ring radii for SU~Aur and RY~Tau correspond to
10 R$_{\star}$ and 11 R$_{\star}$, much larger than the expected
magnetic truncation radius (3 -- 5 R$_{\star}$; \citet{shu94}).  In \S
\ref{scatter} we show that emission from gas between the magnetic
truncation radius and R$_{\rm dust}$ can reconcile accretion disk
models with our observations.

The position angle coverage of the PTI data allow us to constrain the
inclination of infrared emission.  The $\chi^2/{dof}$ improves
substantially for SU Aur and RY~Tau for the inclined models as
compared to the face-on models, but the DR~Tau data do not provide a
good constraint on the inclination given the large error bars and
because the source is at best marginally resolved.  For RY~Tau, our
inclination angle of 19\arcdeg\ -- 25\arcdeg\ agrees with that derived
by \citet{koe95} from resolved millimeter emission (25\arcdeg).
However, the position angle is not well constrained by our data
(98\arcdeg\ $\pm$ 40\arcdeg\ for the uniform disk and 110\arcdeg\
$\pm$ 22\arcdeg\ for the ring) and does not agree with the PA of the
millimeter emission (48\arcdeg\ $\pm$ 5\arcdeg, \citet{koe95};
27\arcdeg\ $\pm$ 7\arcdeg, \citet{kit02}) and is not orthogonal to the
jet PA of 110\arcdeg\ from \citet{sta04}. Our inclination angle of
52\arcdeg $\pm$ 10\arcdeg\ agrees with the 60\arcdeg\ estimate of
\citet{unr04} based on the photometric period and line widths.
\citet{muz03} find high (86\arcdeg) inclination values for both RY Tau
and SU Aur, which are not supported by the PTI data, particularly for
RY~Tau, and are also inconsistent with the low visible extinctions
(A$_v$=2.1 and 0.9, respectively).  At such high viewing angles, the
star would be occulted by the flared circumstellar disk.  If there is
a large incoherent component for any of the sources, then our simple
geometric fits will underestimate the inclination angle as an
incoherent contribution is independent of baseline.  However, in our
detailed models (\S \ref{scatter}) for these three objects, the
extended light contribution is less than 10\%, which is insufficient
to change the measured RY Tau inclination by 60\arcdeg.

\section{Detailed radiation transfer models}
\label{scatter}

One of the major uncertainties in the simple fits presented above
is the assumption of no extended emission within the 1\arcsec\ PTI field
of view.  To address this issue directly, we have calculated
radiative transfer models for SU Aur, DR Tau and RY Tau.
The input properties are given in Table \ref{table:source} and 
the goal is to match the PTI data and the SED.

\subsection{Monte Carlo radiation transfer code}
\label{MonteCarlo}

We use the Monte Carlo radiative equilibrium technique of
\citet{bjo01}, updated by \citet{wal04}, to self-consistently model
each of our target sources. This code iteratively solves for the disk
density structure, assuming the dust and gas are well-mixed with a
standard gas to dust ratio of 100:1 and the system is in vertical
hydrostatic equilibrium. In addition to stellar irradiation, the code
includes accretion and shock/boundary layer luminosity calculations
according to \citet{cal98}. Multiple scattering is
treated alongside the heating and reprocessing of photons in the
disk. Output data can be used to produce synthetic SEDs and
multi-wavelength images for any viewing angle of the disk system. For
more detailed description of the code and its updates see
\citet{woo02a,woo2b,whi03a,whi03b,wal04} and references therein.

\begin{figure}[h!]
\begin{center}
\epsscale{0.9}
\plotone{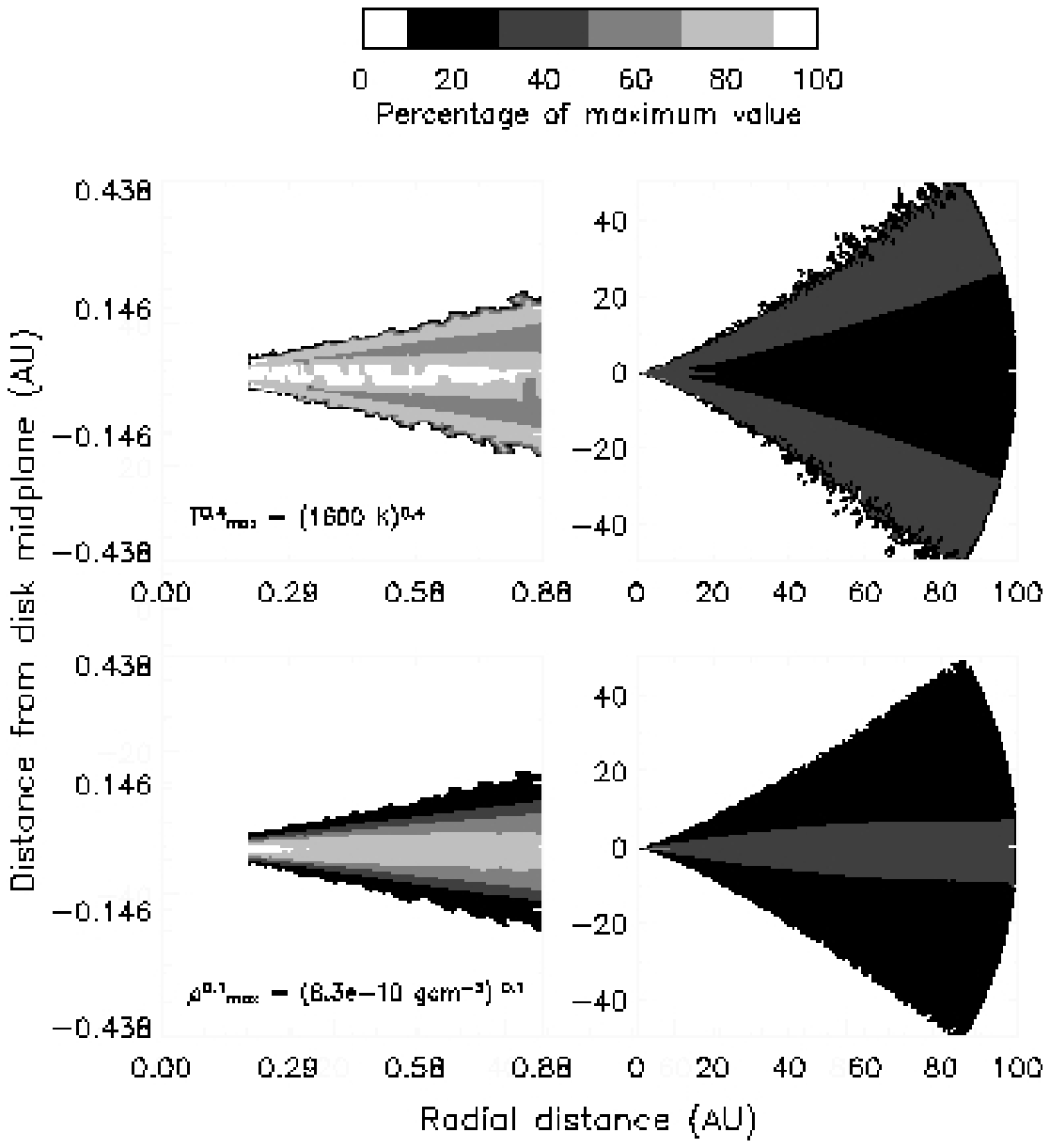}
\caption{Temperature and density distributions for
an example disk model.  The upper images are temperature scaled
to the 0.5 power and the lower are density to the 0.1 power. Note
the geometrically thin gas is not shown.
\label{density:plot}
}
\end{center}
\end{figure}

The code computes the flared density structure of a steady accretion
disk extending from the inner dust destruction radius to a specified
outer radius (Figure \ref{density:plot}).  The Monte Carlo technique
naturally accounts for radiation transfer effects and the heating and
hydrostatic structure of the inner wall of the dust disk.  The
vertical height of the inner wall of dust is not preset, but rather
calculated as part of the modeling process.  For these models the
scale height of the density distribution is 0.3 to 0.7 R$_{\star}$ at
the inner edge.  The position of the inner dust disk edge, R$_{\rm
dust}$ is determined from the destruction temperature of silicates,
taken to be 1600~K \citep{dus96}.  Within the disk we adopt the
dust-size distribution used for the modeling of HH30 IRS and GM Aur
\citep{woo02a,sch03,ric03}.  With a distribution of grain sizes or
compositions, the dust destruction may take place over a range of
radii, but this is beyond the scope of our work.  \citet{mon02}
discuss the constraints on the dust properties from infrared
interferometry observations.

In order to match the new PTI observations, R$_{\rm dust}$ for some
sources was large enough ($>$ 0.2 AU) that continuum emission from gas
within R$_{\rm dust}$ becomes significant. The structure and
temperature of the gas disk is not computed self-consistently in our
models.  Instead, accretion luminosity is emitted following the
temperature structure of an optically thick accretion disk 
\begin{equation}
T_{gas}(R) = (\frac{3 G M_{\star} \dot{M}}{8 \pi \sigma R^3})(1 - \sqrt{R_{\star}/R})^{1/2})^{1/4}
\end{equation}
(e.g.,
\citet{lyn74, pri81}) where $R$ is the radial distance in the mid-plane.
The gas disk is assumed to be
infinitely thin, so after being emitted, the ``accretion photons'' do
not encounter any opacity in the gas, but may be scattered and
absorbed, and produce heating in the dust disk.  Clearly this is a
simplification for the gas emission, but is sufficient for our models.
The assumed geometry of the gas disk is supported by recent modeling
by \citet{muz04} of Herbig Ae/Be sources in which the gas disk is
geometrically thin, allowing direct radiation of the inner dust disk.
Future work will investigate the effects of possible shielding of the
dust disk by a flared and possibly optically thick, inner gas disk.

The gas disk extends down to the magnetic truncation radius (R$_{\rm
gas}$) at which point, material is thought to be channeled along
magnetic field lines onto the star at a high latitude shock zone
(e.g. \citet{dal03}, \citet{ken94}).  We assume the gas disk is
truncated at a magnetospheric radius dependent on the stellar radius,
mass, accretion rate and surface magnetic field \cite{gho79}.  For DR
Tau and RY Tau we assume kiloGauss magnetic fields and truncate the
gas disk at 5 R$_{\star}$.  For SU Aur, thought to be more weakly
magnetic and with inconclusive evidence for hot spots \citep{unr04} we
use 2 R$_{\star}$ for R$_{\rm gas}$.  We assume photons emitted from the
shock/boundary layers have a spectrum of an 8000~K Planck function
\citep{cal98} and are emitted along with stellar photons as in
\citet{muz03}.  

For each model, the stellar luminosity has been fixed as detailed in
Table 1 and for input stellar spectra we use the appropriate Kurucz
(1994) model atmosphere. We also used fixed stellar masses of 2.25
M$_{\odot}$, 2M$_{\odot}$, and 1 M$_{\odot}$ for SU Aur, RY Tau and DR
Tau, respectively \citep{coh79,ken94}; note that the stellar mass is
not a critical parameter in the near-infrared.  The disk properties
such as mass, accretion rate and inclination were varied in order to
produce a grid of synthetic SEDs. These models allowed us to explore
likely parameter configurations.

\subsection{Visibility calculation}

\begin{figure}[h!]
\begin{center}
\epsscale{0.5}
\plotone{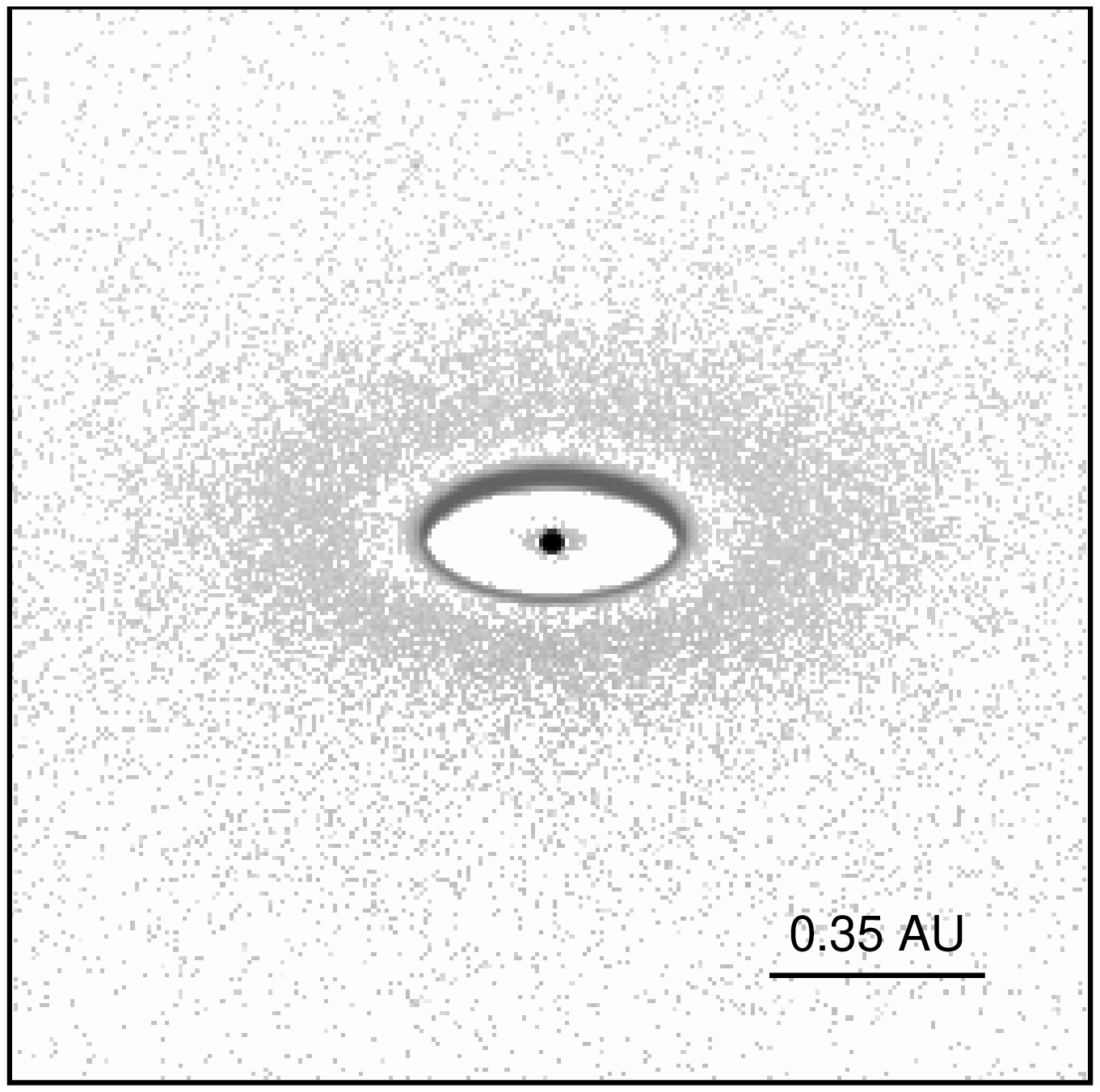}
\plotone{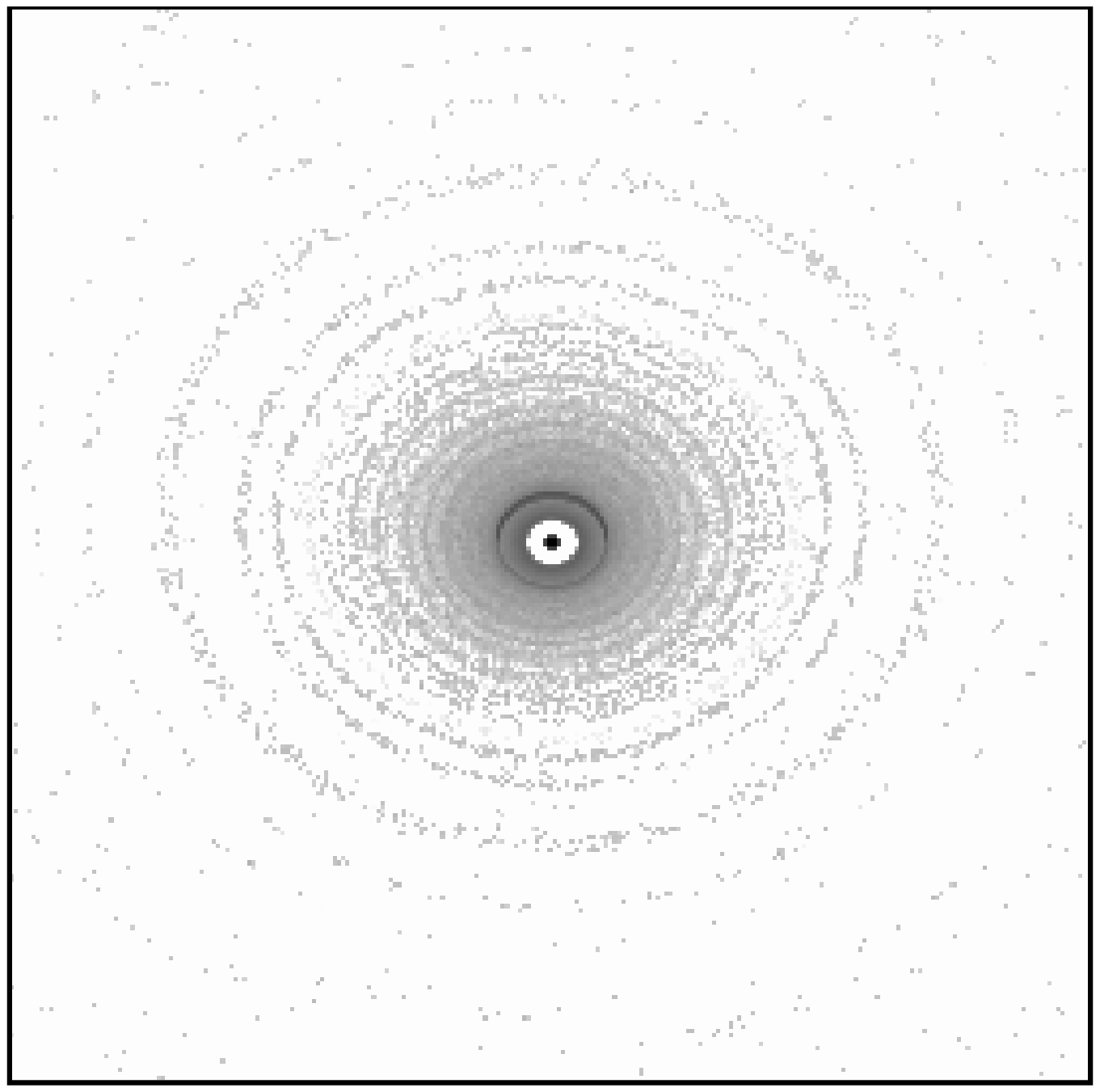}
\plotone{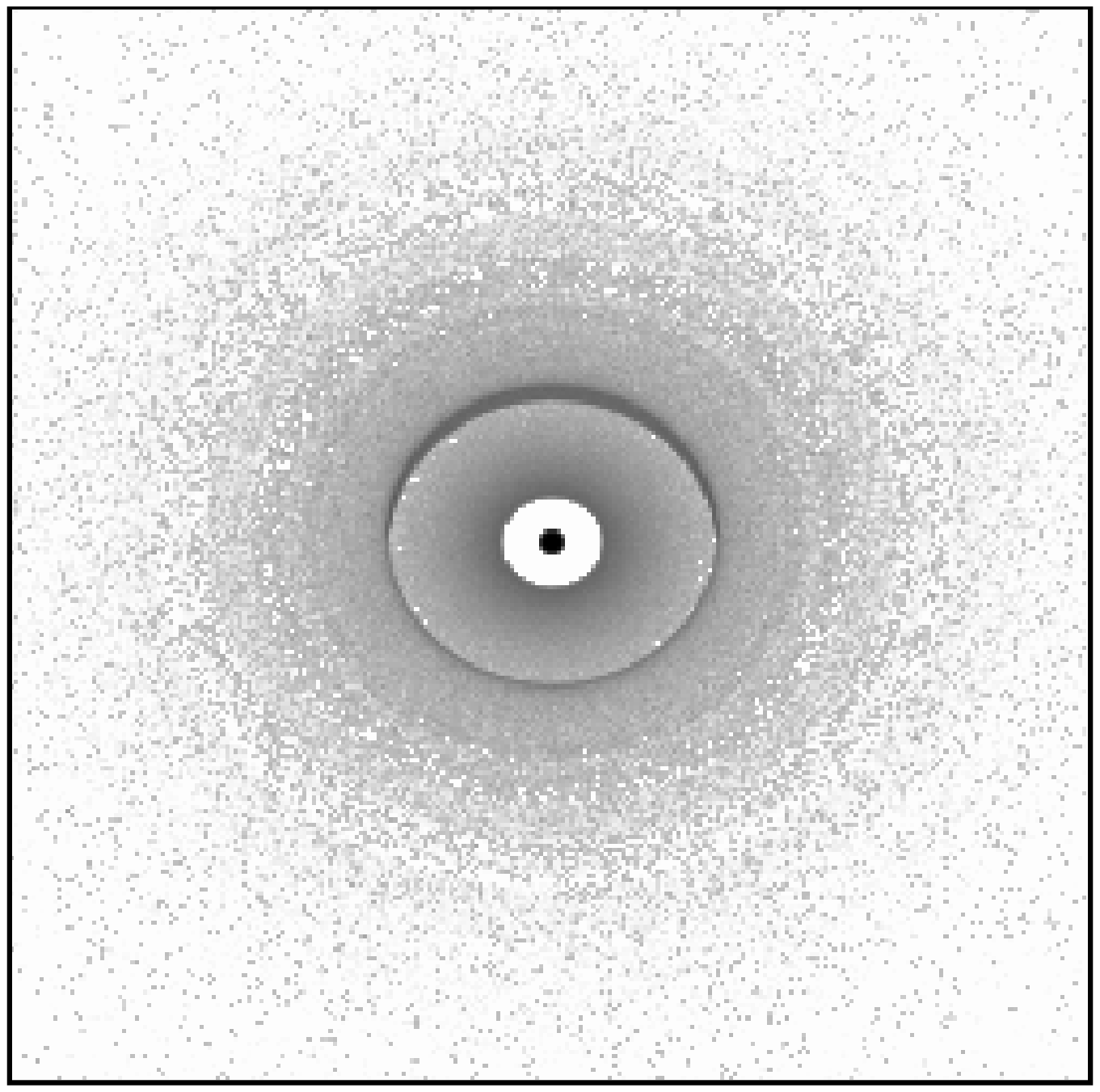}
\caption{Model images for SU Aur (top), DR Tau (middle) and RY Tau
(bottom).  The flux has been scaled to the 0.15 power to provide
better contrast in the image.  Each image is 12.5 milliarcsec or 1.75 AU
across.  For comparison, all models are shown with the same position
angle.  
\label{model:plot}
}
\end{center}
\end{figure}

To calculate model visibilities, a simulated K band image was created
using the Monte Carlo models with pixel size 0.05 mas and a width of
12.5 mas.  The pixel size was chosen to be much smaller than the
fringe spacing of 4 mas and the total size was a compromise between
calculation time (large images are computationally intensive) and
capturing the relevant structure.  The outer size is large enough to
contain any component which would contribute substantially to the
model visibility.  For example, a thin ring with an inner radius of 3
mas has V$^2$=0.01 on the shortest PTI baseline.
The K band emission in the models is dominated by structures a few mas
in size or less (Figure \ref{model:plot}).  As discussed in \S
\ref{model}, any emission within the 1\arcsec\ FOV will contribute
incoherently.  To calculate the extended component in the model, a larger
image is also constructed with 2 mas pixels and a 1\arcsec\ field.
The emission outside the central 12 mas is calculated and included in
the visibility calculation as an incoherent contribution.  The effects
of the 1\arcsec\ Gaussian field of view and the finite fringe envelope
are also included in the visibility calculation.  For each PTI
baseline, the model visibility, including the incoherent flux, was
calculated for the average baseline length and position angle using
the Fourier Transform of the image, assuming a position angle for the
disk as given in \S \ref{model}.  We thus ``observe'' the models as
they would be at PTI.  

\begin{figure}[h!]
\begin{center}
\epsscale{0.6}
\plotone{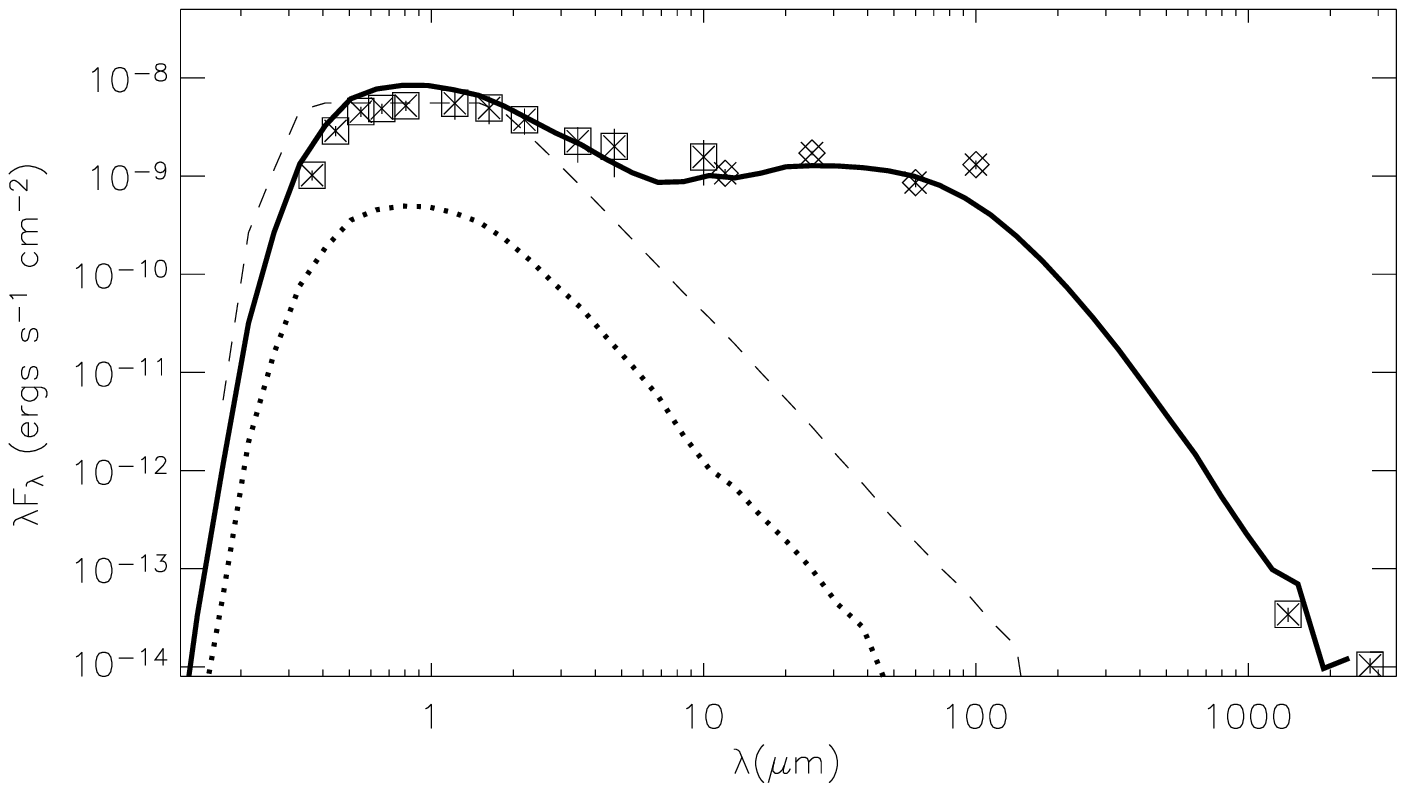}
\plotone{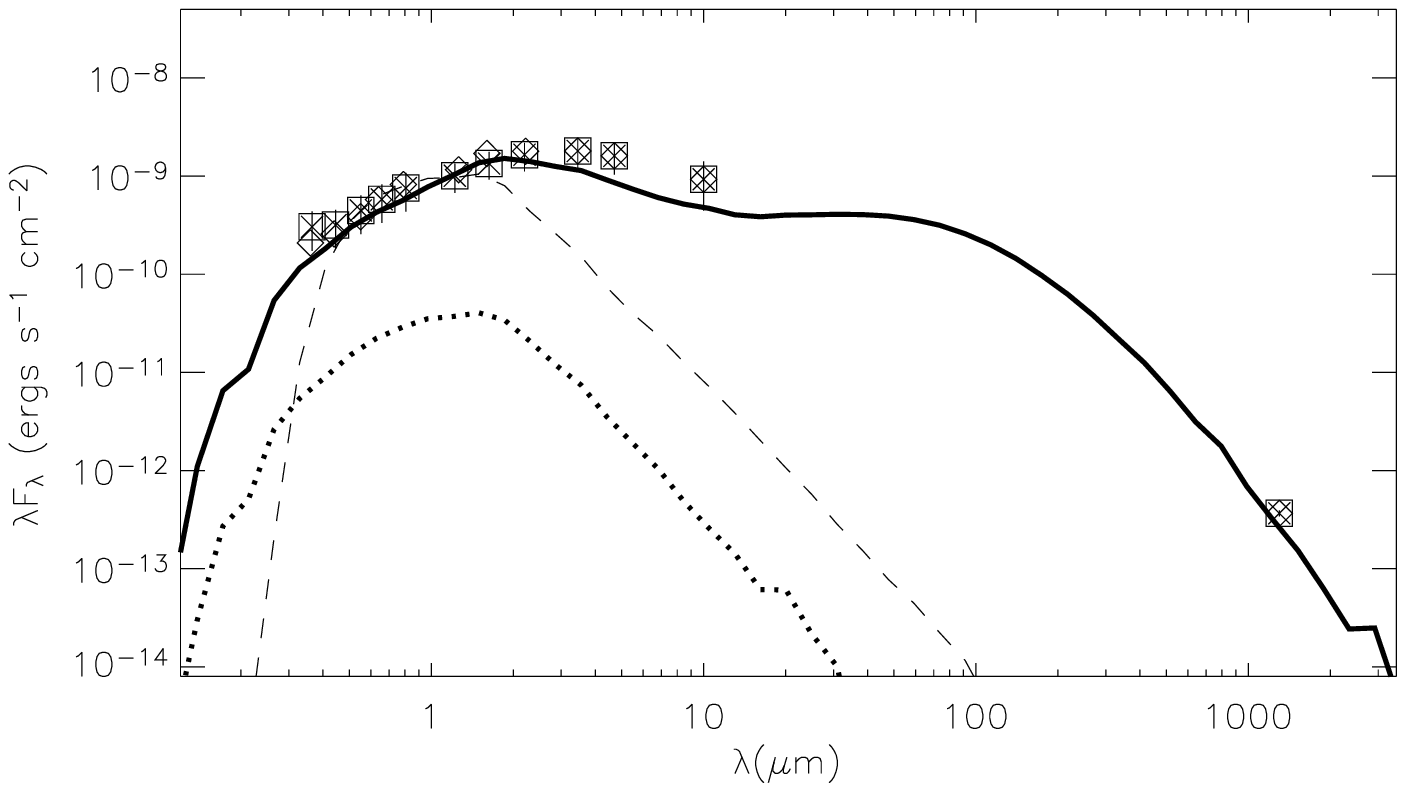}
\plotone{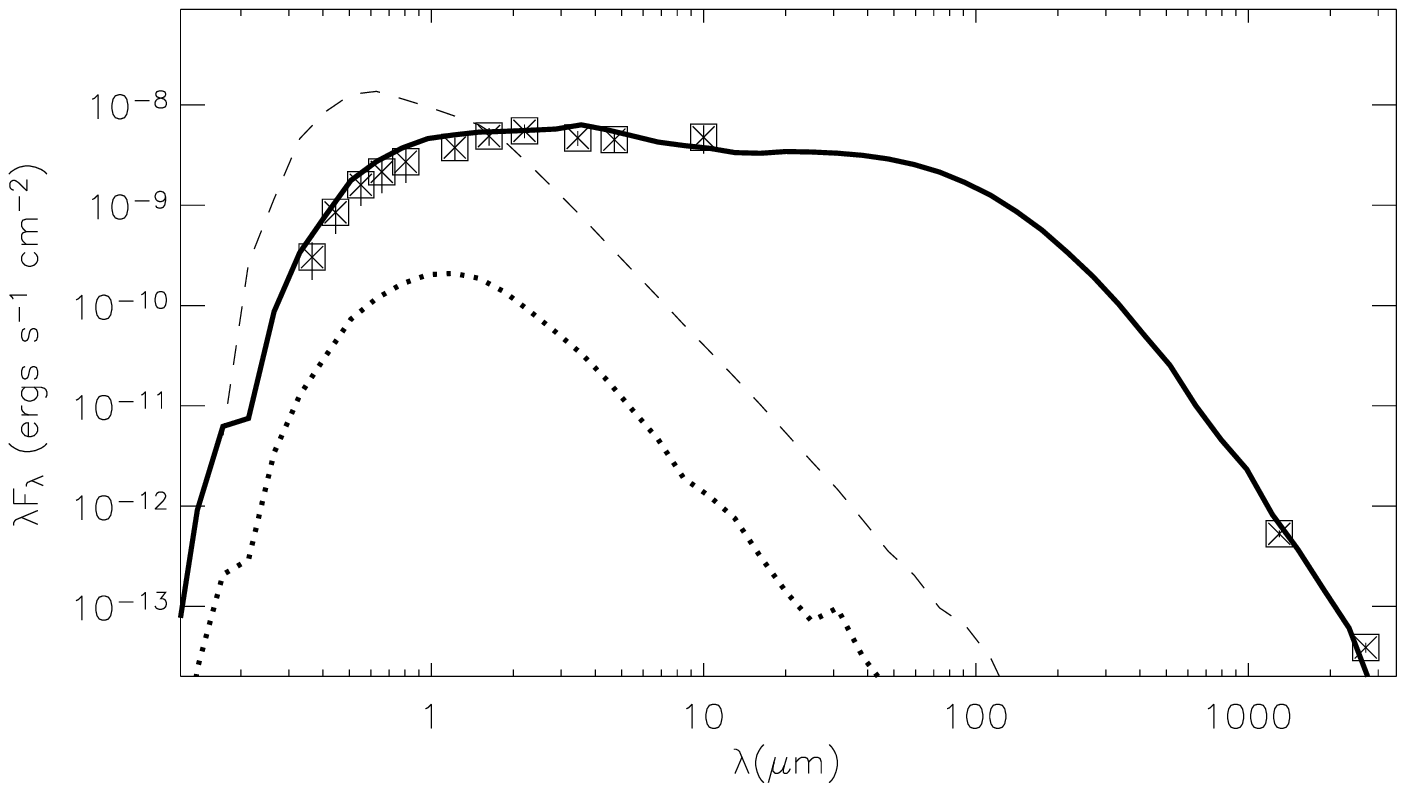}
\caption{SED plots for best-fit models for SU Aur (top), DR Tau (middle)
and RY Tau (bottom).  The model total flux is given by the
solid line, the input stellar spectrum by the dashed line and the
scattering by the dotted line.  Data from \citet{ken95} are given by squares and from \citet{eir02} by circles.
\label{fig:SED}}
\end{center}
\end{figure}

The model visibilities and full SEDs were then
compared to the data presented in \S \ref{observations} and SED data
taken from the literature. The optical and infrared SED data are taken
from the compilation of \citet{ken95} and are not contemporaneous but
instead represent an average for each source and the millimeter data
are taken from \cite{ake02} and \cite{bec90}.  For each object a set of models
was calculated to explore the disk parameters to find a viable model
and an example selection of models for each object is given in Table
\ref{model:table}.  The parameter space chosen for the inclination
angle was restricted using the results from the geometric fits.  The
model with the total lowest $\chi^2$ is shown for each object in Fig
\ref{model:plot} and the corresponding SED fit in Fig \ref{fig:SED}.

\begin{table}
\begin{center}
\begin{tabular}{llllllllll}
Model & $\dot{M}$ & r$_{in}$  & $M_{disk}$ & incl & Lacc+shk & $\chi^2_{PTI}$ & $\chi^2_{SED}$ & $\chi^2_{total}$ & notes \\ 
& M$_{\odot}$/yr & AU & M$_{\odot}$ & deg & L$_{\odot}$ \\      \hline
\multicolumn{10}{c}{SU Aur} \\ \hline
SU-A & $1 \times 10^{-9}$ & 0.21 & 0.001 & 60 & 0.02 & 7 & 118 & 125 \\
SU-B &  $1 \times 10^{-9}$ & 0.21 & 0.001 & 50 & 0.02 & 29 & 101 & 130 \\
SU-C &  $1 \times 10^{-9}$ & 0.22 & 0.001 & 50 & 0.02 & 60 & 198 & 258 \\
SU-D &  $2 \times 10^{-9}$ & 0.21 & 0.001 & 50 & 0.02 & 48 & 163 & 211 \\
SU-E &  $4 \times 10^{-9}$ & 0.24 & 0.005 & 50 & 0.02 & 99 & 893 & 992 \\ \hline
\multicolumn{10}{c}{DR Tau} \\ \hline
DR-A &  $8 \times 10^{-8}$ & 0.09 & 0.16 & 30 & 1.3 & 12 & 24 & 36 & R$_{gas} = 2 R_{\star}$\\ 
DR-B & $8 \times 10^{-8}$ & 0.09 & 0.16 & 30 & 1.3 & 30 & 19 & 49 \\ 
DR-C & $8 \times 10^{-8}$ & 0.09 & 0.12 & 30 & 1.3 & 30 & 59 & 89 \\
DR-D & $6 \times 10^{-8}$ & 0.09 & 0.15 & 60 & 0.97 & 9 & 55 & 63 \\
DR-E & 0 & 0.08 & 0.08  & 60 & 0 & 3 & 82 & 85 & no accretion\\ \hline
\multicolumn{10}{c}{RY Tau} \\ \hline
RY-A & $2.5 \times 10^{-7}$ & 0.27 & 0.015 & 25 & 4.28 & 8 & 37 & 45 \\
RY-B & $3 \times 10^{-7}$ & 0.27 & 0.015 & 25 & 5.13 & 1 & 294 & 295 \\
RY-C & $2 \times 10^{-7}$ & 0.27 & 0.012 & 25 & 3.42 & 135 & 31 & 166 \\
RY-D & $2.5 \times 10^{-7}$ & 0.27 & 0.015 & 25 & 4.28 & 4 & 145 & 149 & no gas \\
RY-E & $2.5 \times 10^{-7}$ & 0.27 & 0.015 & 25 & 4.28 & 36 & 313 & 349 & envelope \\ \hline
\end{tabular}
\caption{Representative model parameters for each source.  The best
fit model is listed first. $\chi^2$ values given have {\it not} been
normalized by the degrees of freedom. 
\label{model:table}}
\end{center}
\end{table}

\subsection{Other emission components}
\label{extended}

Before discussing the results of the Monte Carlo models, we discuss
other possible physical components which have not been included in our
models.  As discussed in \S \ref{model}, a binary companion will
contribute coherently or incoherently to the measured visibility
depending on the separation from the primary.  RY Tau was classified
as a ``Variability induced mover'' from Hipparcos data and
\cite{ber89} found a solution in which the possible companion had a
minimum separation of 24 mas.  However, the K band speckle
interferometry survey of \citet{lei93} did not detect a companion for
RY Tau in the angular range of 0\farcs1 to 10\arcsec\ and HST archival
images of RY Tau from WFPC2 show only a single point source.  For the
incoherent contribution from a companion to account entirely for the
measured visibility the K band flux ratio would be 0.81 to 1.44
(secondary/primary), but this could not account for the change in
visibility with baseline length and orientation.

Another likely contributor of infrared emission is a circumstellar
envelope.  An envelope can be a source of scattered and thermally
reprocessed starlight and can also veil emission from the central star
and accretion disk (see e.g. the models of \citet{cal97};
\cite{whi03b}).  However, SU Aur, DR Tau and RY Tau are all Class II T
Tauri stars and have visual extinctions $\lesssim$ 2.  In general, Class
II sources are thought to have little or no envelope remaining (see
e.g. \citet{mun00}).  RY Tau shows near-infrared CO lines in
absorption \citep{naj03} which \citet{cal97} cite as evidence of no
substantial envelope for other Class II sources.

\begin{figure}[h!]
\begin{center}
\plotone{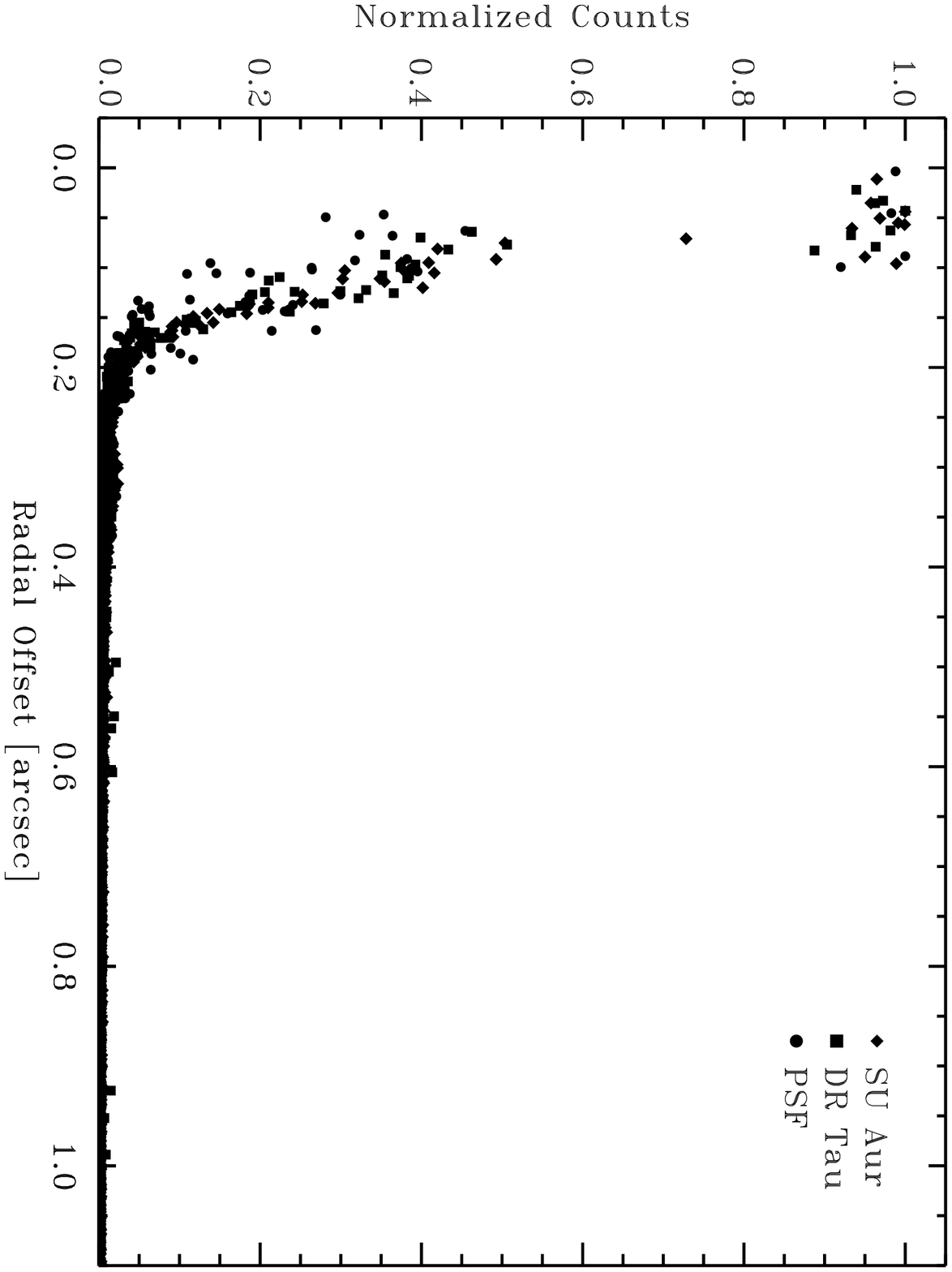}
\caption{Azimuthal averages from HST archival images for SU Aur
(diamonds), DR Tau (squares) and a PSF standard (circles), all using
the F814W filter.  For comparison, each source is self-scaled to the
peak brightness.
\label{psf}}
\end{center}
\end{figure}

To assess the possible presence of emission within the 1\arcsec\ PTI
FOV, we examined HST archival images from the standard imaging
pipeline.  For each source, we found WFPC2 images taken with the F814W
filter on the Planetary Camera CCD.  No extended emission or
additional sources were apparent in the images, however they are
dominated by the central point source.  Azimuthal brightness averages
were computed for comparison to a star extracted from the PSF archive.
The core of RY Tau was saturated to such an extent that we were unable
to find a matching saturated PSF for comparison.  Although the scatter
in the PSF averages do not allow a precise comparison, especially in
the core, SU Aur and DR Tau are dominated by a central source (Figure
\ref{psf}).  This is in contrast with the images of Class I sources,
such as the sample imaged by \citet{pad99} which show images dominated
by scattered light from circumstellar material hundreds of AU in
extent.  Coronagraphic techniques have revealed
some extended emission in RY Tau \citep{nak95} and SU Aur
\citep{cha04} but as detailed in the next section, this emission is
unlikely to contribute substantially in the near-infrared.  We have
also calculated an example model with a disk and an envelope for RY
Tau (\S \ref{rytau}).

\subsection{Results}
\label{results}

As the wavelength range used for the SED comparison ranges from 0.365
microns to 3 millimeters, the model SED is sensitive to many model
parameters, from the extinction to the outer disk size and mass.  As
expected, the infrared visibility is sensitive to only a few model
parameters, particularly the inner radius, the inclination angle and
the luminosity.  Each object is considered separately below, but
the general conclusion is that these models, which include the
contribution from extended emission, support the simple geometric
models in the large value of R$_{\rm dust}$ found for RY Tau and SU Aur.
In part this is because the models contain incoherent
contributions at 2 microns (here defined as flux from scales greater
than 10 mas) which were less than 6\% for all sources, and could still
reproduce both the SED and the infrared interferometry observations. 
High resolution infrared imaging observations would further constrain the
extended emission component of these models.  The K band excess flux
from the models is also close to the veiling values used in \S
\ref{model}, with F$_{\rm excess}$/F$_{\rm total}$ values of 0.4, 0.74
and 0.68 respectively for SU Aur, DR Tau and RY Tau.

\begin{figure}[h!]
\begin{center}
\epsscale{0.6}
\plotone{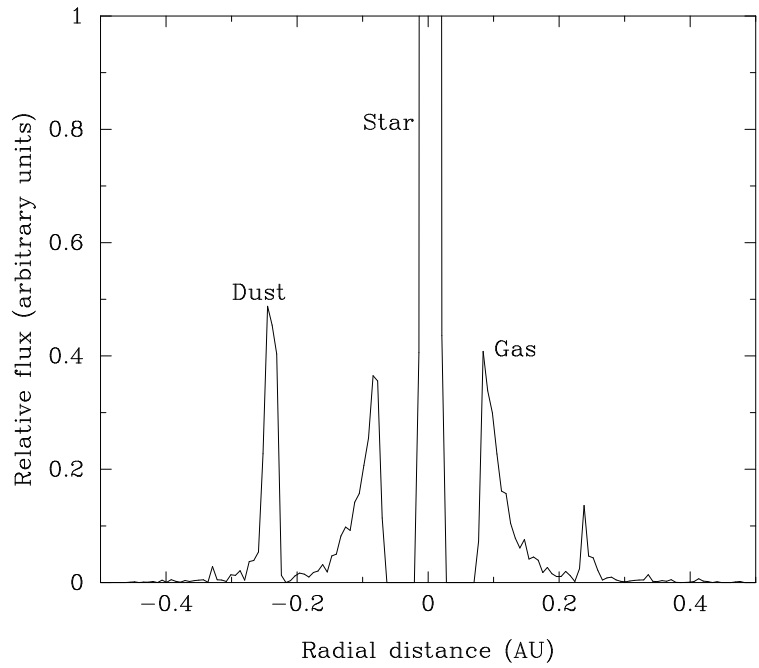}
\caption{A cross section of the RY Tau model through the center of
the source showing the relative flux contributions of the gas and
dust emission.  The slice is oriented such that the inner dust wall
facing the observer is on the left.
\label{fig:slice}}
\end{center}
\end{figure}

The second general conclusion is that emission from gas within R$_{\rm
dust}$ is a significant component of the near-infrared emission if
R$_{\rm dust}$ is large.  For our three objects this is most evident
in the RY Tau model.  The relative flux of the gas and dust components
for RY Tau can be seen in Figure \ref{fig:slice}, which shows a cut
through the model image with the inner dust wall facing the observer
on the left in the plot.  In comparison, for the DR Tau model a
smaller R$_{\rm dust}$ is necessary to match the high visibilities
measured at PTI and so R$_{\rm dust}$ and R$_{\rm gas}$ are similar.
For DR Tau, a smaller value of R$_{\rm gas}$ than estimated from the
stellar properties (2 R$_{\star}$ instead of 5 R$_{\star}$) was
necessary to match the data.  We have not explored the value of
R$_{\rm gas}$ extensively in these models, so these values should be
taken as approximate.  Observations by \citet{naj03} of CO fundamental
emission for several single T Tauri stars similar to our targets
(e.g. BP Tau) found the inner CO radius to be smaller than the
calculated corotation radius for 5 out of 6 sources with CO inner
radii of 0.02 to 0.09 AU.  Modeling of Herbig Ae/Be sources by
\citet{muz04} found emission from the inner gas exceeded the stellar
emission for accretion rates $> 10^{-7}$ M$_{\odot}$/yr.

To confirm the effect of the gas emission, a model was constructed for
RY Tau in which the gas was artificially removed from the region
within R$_{\rm dust}$, which was set to match the measured PTI
visibilities.  However, the SED fit for this model (Table
\ref{model:table}, model RY-D) is not as good as for the model with
gas emission.

\subsubsection{SU Aur}

For SU Aur, both the SED and the measured visibilities are well fit by
model SU-A.  Figure \ref{model:plot} shows the inner region of this
model, with the star and the inner dust disk edge producing the K band
emission.  At an inclination angle of 60\arcdeg\ the inner edge of the
dust disk facing the observer is clearly brighter.  We find R$_{\rm dust}$
= 0.21 AU, similar to the inclined ring model radius of
0.18 AU from Table \ref{table:fits}.  The gas emission is visible close
to the star (R$_{gas} = 2$R$_{\star}$) but the emission is limited by
the small surface area at this radius.

Recent observations by \citet{cha04} have traced extended emission at
K out to radii of 2\farcs6.  This study did not measure the scattered
light within 1\arcsec\ so does not help constrain the PTI data, but
does suggest that a complete model for SU Aur would include an
extended scattered component; however, their measured K flux from
1\arcsec\ to 2\farcs6 was only 4\% of the 2MASS K flux, so neglecting
this component adds only a small error to the SED fit.

\subsubsection{DR Tau}

It was not possible to fit both the PTI data and the SED data well
with these models.  The best-fit model listed in Table
\ref{model:table} underestimates the infrared visibility and
underestimates the SED throughout the near and mid-infrared (Figure
\ref{fig:SED}).  In order to produce an inner disk radius small enough
to fit the PTI data, the model must contain no accretion (model DR-E), which
drastically underestimates the SED and disagrees with the accretion
diagnostics observed for the source \citep{ken94,muz03}.  The infrared
photometry (Table \ref{IRphot:table}) does not reveal substantial
variations recently, however the optical veiling for DR Tau has been highly
variable \citep{gul00} and as the SED data and PTI measurements are
not contemporaneous, there may be issues with source variability in
our modeling.  Also note that the models in Table \ref{model:table} have
a lower inclination ($30 \arcdeg$) than given by the geometric fits
($60 \arcdeg \pm 30 \arcdeg$) but within the uncertainty.

We compared the photometry from \citet{ken95} which
are averages of measurements from the literature to the optical and
infrared contemporaneous SED from \citet{eir02} and the main deviation
is slightly lower fluxes at $u,b,v$ for the contemporaneous SED, which
does not improve our model fits.  In the variability study by
\citet{skr96} DR Tau showed no trend in color with brightness changes,
suggesting the variability is not due to large extinction changes.
The variability of DR Tau has been modeled as a hot spot on the
stellar photosphere \cite{ken94}, but the models here do not attempt
to model DR Tau with that level of detail.

\subsubsection{RY Tau}
\label{rytau}

RY Tau has the largest R$_{\rm dust}$ of the three sources and gas
within R$_{\rm dust}$ contributes substantially to the infrared
emission.  This second component in the disk emission means that the
simple ring model is an {\it underestimate} of R$_{\rm dust}$.
Although this same gas component is present in the models for all
three sources, the contribution to the K band flux is largest for
RY~Tau as R$_{\rm dust}$ is larger than for the other two sources.
For RY Tau, the best fit model was relatively close to the optimal
parameters for both the PTI and SED data.  The PTI data are best
modeled by a higher total luminosity and accretion rate (model RY-B)
than the SED data.

A reflection nebulosity has been observed extending to
$\sim$40\arcsec\ from RY~Tau at visible wavelengths.  However, this is
unlikely to contribute substantially at K as the reflection
component is only 2\% of the total flux at 0.9 $\mu$m and scattering
decreases with increasing wavelength.  To test the effect of an
envelope on the predicted visibilities and SED, a model was calculated
using the disk properties of model RY-A with an envelope using the
same gas and dust radii, 0.01 times the disk mass and an infall rate
of 1 $\times 10^{-7}$ M$_{\odot}$/yr for the envelope.  As seen in
Table \ref{model:table}, this model does not fit either the SED or the
visibilities as well.  Other models with higher envelope masses were
also calculated and had even worse fits to the SED.  It may be
possible to better match the SED with a different disk and envelope
combination, however we found no observational evidence for 
substantial near-infrared emission from an envelope.

We note that \cite{cal04} have recently characterized RY Tau as a G1
star, substantially earlier than previous spectral type
determinations.  However, their stellar properties (R$_{\star}$ = 2.9
R$_{\odot}$, M$_{\star}$ = 2.0 M$_{\odot}$) agree reasonably well with
the values we used (Table \ref{table:source}).  We used a slightly
lower effective temperature (5782 K compared to 5945 K) and a higher
luminosity (12.8 L$_{\odot}$ compared to 9.6 L$_{\odot}$).  A lower
stellar luminosity would require a more massive disk and higher
accretion to produce the same flux at longer wavelengths, but the
general properties of a large R$_{dust}$ would not change.  The
accretion rate of the model presented here, $2.5 \times 10^{-7}$
M$_{\odot}$/yr, is actually higher than the \citet{cal04} estimate of
$6.4-9.1 \times 10^{-8}$ M$_{\odot}$/yr.

\section{Conclusions}

Infrared interferometric observations of T~Tauri stars are used to
constrain the inner disk properties.  Detailed models were presented
for SU Aur, RY Tau and DR Tau to reproduce both the interferometry
observations and the spectral energy distribution.  For both the
simple geometric fits to the interferometry data and the Monte Carlo
disk models which include accretion and scattering, the inner dust
radius ranges from 0.05 to 0.3 AU.  Extended envelopes were not needed
to reproduce the SED for these sources, although additional high
resolution infrared images would help in constraining the extended
emission on larger scales (tens to hundreds of AU).  However, the
significant variations in the visibility with baseline length and
orientation seen for SU Aur and RY Tau require a resolved component to
be present as an extended component produces a constant visibility
reduction.  Although the models parameters given here may not be a
unique solution to the data set used, they are consistent with the
domination of the near-infrared excess by thermal emission from the
disk, as expected for Class II sources.

The SU Aur model agrees well with the size derived from geometric fits
to the interferometer data alone.  The SED and PTI visibilities for DR
Tau can not both be fit with these models, perhaps due to source
variability.  For RY Tau, the model predicts significant emission in
the K band from gas within the inner dust disk radius.  This gas
emission at infrared wavelengths is generally not considered in the
simple models used to fit interferometric data and when this emission
is present results in an underestimate of the dust radius when using a
simple ring model.

Future work will extend the observational database to study more
sources and probe the innermost regions of disks.  Interferometric
observations at a second wavelength would add additional constraints
on the inner disk structure.  From a theoretical perspective, we are
extending the Monte Carlo codes to include gas opacity in the inner
disk and self-consistently calculate its structure.  The data and
models presented in this paper clearly show that inner gas disks
cannot be ignored and are required when fitting observations that
probe the inner regions of disks.

\acknowledgments

This work was performed at the Michelson Science Center, California
Institute of Technology, under a contract with the National
Aeronautics and Space Administration.  Data were obtained at the
Palomar Observatory using the NASA Palomar Testbed Interferometer.
Science operations with PTI are possible through the support of the
PTI Collaboration({\tt
http://huey.jpl.nasa.gov/palomar/ptimembers.html}) and the efforts of
Kevin Rykoski.  JAE acknowledges support from a Michelson Graduate
Research Fellowship.  This work has made use of software produced by
the Michelson Science Center.  Pomona College would like to
acknowledge the support of the NSF ARI and CCLI grants in providing
funds for development of the infrared camera and the Pomona College
1-meter telescope.  This work has made use of the SIMBAD database,
operated at CDS, Strasbourg, France, and the NASA/IPAC Infrared
Science Archive, operated by the JPL under contract with NASA.  This
work utilizes observations made with the NASA/ESA Hubble Space
Telescope, obtained from the data archive at the Space Telescope
Science Institute. STScI is operated by the Association of
Universities for Research in Astronomy, Inc. under NASA contract NAS
5-26555.

\end{document}